\documentclass[aps,pra,twocolumn,showpacs]{revtex4}

\usepackage[normalem]{ulem}

\input{epsf}

\begin{document}

\title{Zero-Temperature Equation of State and Phase Diagram of Repulsive Fermionic Mixtures}

\author{E. Fratini and S. Pilati}
\affiliation{The Abdus Salam International Centre for Theoretical Physics, 34151 Trieste, Italy}

\date{\today}

\begin{abstract}
We compute the zero-temperature equation of state of a mixture of two fermionic atomic species with repulsive interspecies interactions using second-order perturbation theory.
We vary the interaction strength, the population and the mass imbalance, and we analyze the competition between different states: homogeneous, partially separated and fully separated.  The canonical phase diagrams are determined for various mass ratios, including the experimentally relevant case of the $^6$Li-$^{40}$K mixture. 
We find substantial differences with respect to the equal-mass case: phase separation occurs at weaker interaction strength, and the partially-separated state can be stable even in the limit of a large majority of heavy atoms.
We highlight the effects due to correlations by making comparison with previous mean-field results.
\end{abstract}


\maketitle

\section{Introduction}
\label{Introduction}
Thanks to the possibility to tune the interaction strength using Feshbach resonances~\cite{chin}, atomic gases offer experimentalists a unique venue to investigate the effects of strong correlations in many-fermion systems.  
Two-component Fermi gases with strong attractive interactions, obtained following the lower branch of the resonance, have been addressed in numerous theoretical and experimental studies (see Ref.~\cite{giorgini} for a review).
More recently, attention has also been given to the upper branch, where a metastable atomic gas with repulsive interaction can be realized~\cite{jo09}.
An important open issue is whether ferromagnetic behavior can be induced by increasing the repulsion strength.
This phenomenon was first discussed by Stoner as a paradigm to explain itinerant ferromagnetism in transition metals~\cite{stoner}.
While, according to Stoner's mean-field theory, at zero temperature a ferromagnetic instability takes place when $k_Fa=\pi/2$ ($k_F$ is the Fermi wave-vector and $a$ the $s$-wave scattering length), more accurate many-body theories predict a considerably weaker critical interaction: $k_Fa \simeq 0.8$~\cite{conduit09,PRL2010,chang2011}.
Since, in the general case, all available computational methods for many interacting fermions, including quantum Monte Carlo simulations~\cite{troyer}, must adopt some approximations or unproven assumption, a benchmark against experimental results is of outmost importance.
Early experimental evidences consistent with the putative itinerant ferromagnetic state have been reported in Ref.~\cite{jo09}. However, subsequent theoretical~\cite{pekker} and experimental studies~\cite{ye2012,sanner2012} have demonstrated that, at strong interactions, molecule formation due to three-body recombinations plays the dominant role and induce local heating, thus preventing the observation of the ferromagnetic instability.
Theoretical studies indicate that the reach of ferromagnetism could be favored in several ways, including using narrow Feshbach resonances~\cite{expgrimm,massignanEPJ,massignan}, considering low dimensional and confined configurations~\cite{jochim,blume,zinner,conduit2013,cui}, and by loading shallow optical lattices ~\cite{DFT,PRL2014} or optical-flux lattices~\cite{cooper}.\\
Since the ferromagnetic transition corresponds to the equal-mass limit of the phase separation in a mixture of two components with different masses, it has been proposed to use mixtures of different atomic species as an alternative route to address itinerant ferromagnetism. 
In fact, it has been proven that for any interaction strength it is always possible to induce phase separation by sufficiently increasing the mass imbalance~\cite{ho2013}. 
Tuning the mass ratio might also allow to lower the rate of three-body recombinations~\cite{petrov}.
However, the critical interaction strength for phase separation has been determined so far only using mean-field theories~\cite{conduit2011}.
This calls for a more accurate theoretical analysis, which is the subject of this Article.\\
Mass-imbalanced repulsive Fermi-Fermi mixtures can be realised in experiments performed with two fermionic atomic species, or with one bosonic and one fermionic species if the density of the fermions exceeds that of the bosons and a Feshbach magnetic field is tuned to the regime of strong attraction so that all bosons pair up with fermions forming (fermionic) molecules~\cite{fratini,bertaina}.
It is worth mentioning that Fermi-Fermi mixtures with population and mass imbalance have been considered also in the context of Hubbard-type lattice models, and a variety of magnetic phases has been found~\cite{capone,hofstetter}.\\

In this Article, we employ second-order perturbation theory to determine the zero-temperature equation of state of a repulsive Fermi-Fermi mixture. In Section~\ref{methods} we describe our model Hamiltonian and we develop the formalism for the perturbation theory.
In Section~\ref{resultsEOS}, we discuss  a comparison between our second-order theory, the mean-field approximation and, for the equal-mass case, the third-order expansion and Variational Monte Carlo simulations.
In Section~\ref{resultsPhaseDiag}, we present the canonical phase diagrams for different mass ratios and we make comparison with previous mean-field predictions. For a mass ratio corresponding to the $^6$Li-$^{40}$K mixture, we find qualitatively different phase-separated states compared to the equal mass case.
Section~\ref{conclusions} reports a summary with a discussion of the important role played by beyond mean-field effects.

\section{Formalism}\label{methods}
We study a repulsive Fermi-Fermi mixture, composed of two atomic species with (in general) different populations and masses.
The system is described by the following Hamiltonian:
\begin{eqnarray}\label{Hamiltonian}
H&=&\sum_{i}\frac{\mathbf{p}_i^{2}}{2m_>}+\sum_{j}\frac{\mathbf{p}_j^{2}}{2m_<}+\sum_{i,j}U(\left|\mathbf{r}_i - \mathbf{r}_j\right|)   \,,
\end{eqnarray}
where the indices $i$ and $j$ label, respectively, atoms of the first and of the second species, $\mathbf{p}_i$ and $\mathbf{p}_j$ are the corresponding momenta, $\mathbf{r}_i$ and $\mathbf{r}_j$ the corresponding positions, while $m_>$ and $m_<$ are the masses of the two atomic species. 
The system is characterized by the volume $V$ and by the total number density $n=n_> + n_<$,  given by the sum of the densities of the first and second species, $n_>$ and $n_<$, respectively. In this section, without loss of generality, we assume that $n_> \geq n_<$, so that the majority-minority population imbalance $r=(n_>-n_<)/n$ is in the range $ 0 \leq r \leq 1$.
The two species interact via a short-range repulsive potential. A commonly adopted model is the Fermi-Huang pseudopotential $U(r) = \frac{2\pi \hbar^2a}{m_r}\delta ({\bf r}) \left[r \frac{\partial}{ \partial r} \cdots \right]$~\cite{huang,kersonhuang}, where $m_r=\frac{m_>m_<}{m>+m_<}$ is the reduced mass and $a$ is the $s$-wave scattering length. This model is adequate to describe interactions in ultracold dilute gases, also in the presence of broad Feshbach resonances~\cite{chin}. It represents a zero-range interaction acting in the $s$-wave channel, and the term in the square parenthesis is a regularisation operator that removes ultraviolet divergences. While, for $a > 0$, this pseudopotential also supports a two-body bound state that approximates the shallow Feshbach molecule, we are interested in the scattering state which describes the repulsive state in the metastable upper-branch of the resonance.
Notice that the Hamiltonian~(\ref{Hamiltonian}) does not include intra-species interactions.
The interaction parameter is conveniently cast in dimensionless form as $k_Fa$, where $k_F=(3\pi^2n)^{\frac{1}{3}}$ is the Fermi wave-vector. 
For a given mass ratio between minority and majority species $\tilde m={m_<} / {m_>}$, the zero-temperature equation of state, in units of $E_F=\frac{3}{10}N\frac{k_F^2}{m_>}$, can be expanded in the following form:
\begin{eqnarray}\label{EOS_hom}
\tilde E^{HOM}(r,k_Fa)&=&\frac{1}{2}\left[ (1+r)^{\frac{5}{3}}+\frac{1}{\tilde m}(1-r)^{\frac{5}{3}}\right]+\\\nonumber
&+&\frac{5}{9\pi} k_Fa \frac{1+\tilde m}{\tilde m} (1+r)(1-r) +\tilde E^{(2)}\,.
\end{eqnarray} 
The first and the second terms in the right-hand side represent, respectively, the energy of a noninteracting mixture and the mean-field interaction energy. Beyond mean-field corrections, indicated by $\tilde E^{(2)}$, have been calculated by Kanno~\cite{kanno1,kanno2} for the equal-mass case ${\tilde m} = 1$ - which corresponds to a (possibly polarized) spin-1/2 Fermi gas - by using second order perturbation theory. For the case of balanced populations ($r = 0$), this correction was first determined by Lee, Huang and Yang~\cite{huang,lee}.
While the mean-field term is linear in the interaction parameter $k_Fa$, second-order perturbation theory is correct to second order in $k_Fa$.
In the present work, we generalize the second-order perturbative calculation to the case of different masses.
By extending the formalism of Refs.~\cite{abrikosov} (see section 1.5) and~\cite{kanno2} to the mass-imbalanced case, one obtains the following expression for the second-order term:
\begin{equation}\label{secondE}
\tilde E^{(2)}=\frac{1}{4\pi^2}\left(\frac{1+\tilde m}{\tilde m}\right)^2 (k_Fa)^2I(r,\tilde m)
\end{equation}
where $I(r,\tilde m) = I^+(r,\tilde m)+I^-(r,\tilde m)$ is the sum of the two integrals:
\begin{eqnarray}\label{Irm1}
I^+(r,\tilde m)&=&-\frac{5}{(2\pi)^3}\int_{FS^+}d^3p_i\,\int_{FS^-}d^3p_j\,\\\nonumber
\int_{FS^+}d^3p_k&&\left(p_i^2+\frac{p_j^2}{\tilde m}-p_k^2-\frac{(\mathbf{p}_i+\mathbf{p}_j-\mathbf{p}_k)^2}{\tilde m}\right)^{-1} \, ,\\
I^-(r,\tilde m)&=&-\frac{5}{(2\pi)^3}\int_{FS^+}d^3p_i\,\int_{FS^-}d^3p_j \label{Irm1b}\\
\int_{FS^-}d^3p_k&&\left(p_i^2+\frac{p_j^2}{\tilde m}-\frac{p_k^2}{\tilde m}-(\mathbf{p}_i+\mathbf{p}_j-\mathbf{p}_k)^2\right)^{-1}\,.\nonumber
\end{eqnarray}
The integration domains are the Fermi Sphere of the majority species $FS^+$ , with radius $p_+=(1+r)^\frac{1}{3}$, and the Fermi Sphere of the minority species $FS^-$, with radius $p_-=(1-r)^\frac{1}{3}$.
Equations~(\ref{Irm1}) and (\ref{Irm1b}) are free of ultraviolet divergences~\cite{abrikosov} and do not depend on details of the inter-atomic interactions other than $a$. Hence, the present theory applies to the repulsive state of the Fermi-Huang pseudopotential~\cite{kersonhuang}, and also to other model potentials with finite range, as long as this range is much smaller than $k_F^{-1}$.
By solving analytically the angular integration and applying two changes of variables: $\mathbf{k}=\mathbf{p}_i+\mathbf{p}_j$ and $\mathbf{q}=\mathbf{k}-\mathbf{p}_k$, we get
\begin{eqnarray}\label{Irm2}
I^+(r,\tilde m)&=&5\int_0^{p_+}dp_i\,\int_0^{p_-}dp_j\,\int_0^{p_+}dp_k\,p_ip_jp_k\,\\\nonumber
&&\left(\int^{p_i+p_j+p_k}_{\vert p_i-p_j\vert+p_k}dk \ln{\left| p_i^2+\frac{p_j^2}{\tilde m}-p_k^2-\frac{k^2}{\tilde m}\right|} +\right.\\\nonumber
&-&\left.\int^{p_i+p_j-p_k}_{\vert p_i-p_j\vert-p_k}dk \ln{\left| p_i^2+\frac{p_j^2}{\tilde m}-p_k^2-\frac{k^2}{\tilde m}\right|}\right)\\\label{Irm2b}
I^-(r,\tilde m)&=&5\int_0^{p_+}dp_i\,\int_0^{p_-}dp_j\,\int_0^{p_-}dp_k\,p_ip_jp_k\,\\\nonumber
&&\left(\int^{p_i+p_j+p_k}_{\vert p_i-p_j\vert+p_k}dk \ln{\left| p_i^2+\frac{p_j^2}{\tilde m}-\frac{p_k^2}{\tilde m}-k^2\right|} +\right.\\\nonumber
&-&\left.\int^{p_i+p_j-p_k}_{\vert p_i-p_j\vert-p_k}dk \ln{\left| p_i^2+\frac{p_j^2}{\tilde m}-\frac{p_k^2}{\tilde m}-k^2\right|}\right)\,.
\end{eqnarray}
To evaluate the integrals in equations (\ref{Irm2}) and (\ref{Irm2b}), we implemented an efficient stochastic integration procedure based on Monte Carlo sampling, allowing us to determine the function $I(r,\tilde m)$ on a dense grid of values of the population imbalance $r$. In the Appendix~\ref{appendix}, we report an accurate parametrization of the data based on a simple empirical functional form for a few relevant cases of mass imbalance, namely $\tilde m= 2,1/2,3/2,2/3,40/6,6/40$. In Section~\ref{resultsPhaseDiag} we employ this parametrization to determine the canonical zero-temperature phase diagrams in the absence of external potentials. By using the local-density approximation, one could extend this calculation to harmonically confined configurations.
%

%%%%%%%%%%%%%%%%%%%%%%%%%%%%%%%%%%%%%%%%%%%%%%%%%%%%%%%%%%%%%%%
\section{Equation of state}\label{resultsEOS}
\begin{figure}[h!]
\epsfxsize=8cm
\epsfbox{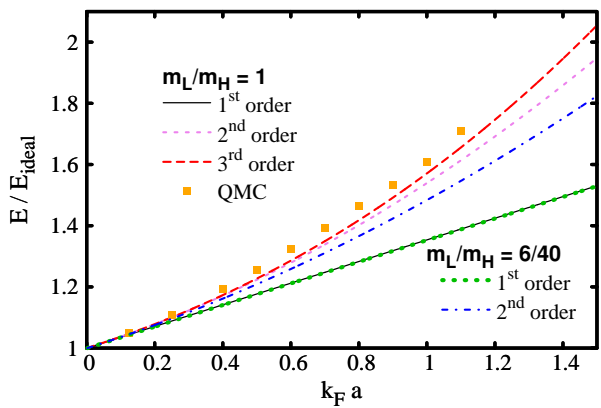}
\caption{Ground-state energy of population-balanced mixtures as a function of the interaction parameter $k_Fa$. The two atomic species have equal masses or the mass ratio $6/40$ corresponding to $^6$Li and $^{40}$K. Analytic expansions valid up to first, second~\cite{huang,lee}, and third order~\cite{efimov,bishop} in $k_Fa$ are compared with the Variational Monte Carlo results~\cite{PRL2010}. The unit is the energy of the noninteracting mixture $E_{\text{ideal}}$. Notice that the mean-field results for the two mass ratios coincide.} 
\label{comparison-EOS}
\end{figure}

In the case of the Fermi-Fermi mixture with equal masses (which can be mapped to a spin-1/2 Fermi gas) the zero-temperature equation of state has been determined using various quantum many-body techniques, including: second-order perturbation theory~\cite{huang,lee,kanno1,kanno2,Duine2005}, perturbative expansions in terms of effective in-medium scattering matrices (correct up to third order in the interaction parameter)~\cite{dedominicis,efimov,bishop,boronat}, diagrammatic theory within ladder approximation~\cite{huang2012}, and variational Quantum Monte Carlo simulations~\cite{PRL2010,chang2011}. The terms of the equation of state beyond the second-order expansion can also be sensible to details of the inter-atomic potential beyond the $s$-wave scattering length $a$, the most relevant being the $s$-wave effective range $r_0$ and the $p$-wave scattering length $a_p$~\cite{bishop}. A detailed analysis of the contribution due to these details was given in Refs.~\cite{bishop,PRL2010,chang2011,boronat}. In Fig.~\ref{comparison-EOS}, we show a comparison between different theories. In the third-order expansion we set $r_0 = 0$ and $a_p = 0$, and include only the high-order terms which depend on $a$. The quantum Monte Carlo simulations of Refs.~\cite{PRL2010,chang2011} were performed using resonant model potentials (with negligible values of $r_0$ and $a_p$) designed to mimic the effect of the Fermi-Huang pseudopotential~\cite{noteQMC}.
We notice that second-order perturbation theory gives a significant contribution beyond the mean-field approximation. The third-order term adds a smaller correction, and provides a result which is quite close to the variational quantum Monte Carlo prediction.
In the case of mixtures with different masses, beyond mean-field contributions to the equation of state have not been determined in previous works. Only the case of a single impurity interacting with an ideal Fermi sea has been studied using diagrammatic ladder approximation~\cite{massignanEPJ,massignan}. In Fig.~\ref{comparison-EOS}, we compare the mean-field prediction with our second-order result for a mixture of two species with equal densities and mass imbalance $m_L / m_H = 6 /40$ (corresponding to the $^6$Li-$^{40}$K mixture). Notice that, when expressed in units of the energy of a noninteracting mixture, the mean-field prediction coincides with the one of the equal-mass case. The second order term adds an important contribution, only slightly smaller than in the equal-mass case.\\
In Section~\ref{resultsPhaseDiag}, we employ our second-order equation of state to determine the zero-temperature phase diagram for various Fermi-Fermi mixtures with different mass imbalances. Since, in the equal-mass case, the equation of state and the phase diagram obtained with the second-order theory~\cite{Duine2005} are close to the result of quantum Monte Carlo simulations~\cite{conduit09,PRL2010,chang2011} and of diagrammatic ladder theory~\cite{huang2012}, we argue that also in the mass-imbalanced case the second-order perturbation theory provides important information about the role of beyond mean-fields effects.

\section{Phase diagrams}\label{resultsPhaseDiag}
When the strength of the repulsive interaction increases, the Fermi-Fermi mixture might become unstable against phase separation into two domains with different local densities and population imbalances. In the case of species with equal masses, this phase separation corresponds to a quantum phase transition to a ferromagnetic state.
It has been proven that, for any fixed interaction strength, such quantum phase transition can always be induced if the ratio of the masses of the two species is increased sufficiently~\cite{ho2013}. It this section, we calculate the critical interaction strength where this phase transition takes place within second-order perturbation theory. For three relevant mass ratios, we determine the zero-temperature canonical phase diagram as a function of interaction strength and global population imbalance. We consider different kinds of phase separation, including the partially-separated state where at least one domain contains atoms of the two species, and the fully-separated state where both domains contain atoms of one species only.\\
We study a system without external potentials at zero temperature, with a fixed (global) number-density of the heavier species (larger atomic mass), indicated as $n_H$, and of the lighter species, indicated as $n_L$. The total volume $V$ is also fixed. In this section, we define the heavy-light population imbalance $r=(n_H-n_L)/(n_H+n_L)$, which will be positive when the heavier atoms are the majority ones, and negative otherwise. The light-heavy mass ratio is $0< m_L /m_H\le1$.
The energy of the separated state is determined by the following expression: 
\begin{eqnarray}\label{EOS_S}
E^{SEP}(r,k_Fa) && =\\\nonumber
\min_{\left\{v_1,n_{H_1},n_{L_1}\right\}} && \left[\right.v_1E^{HOM}(\left|r_1\right|,k_{F_1}a)+\\\nonumber
&&\,\, \left. v_2E^{HOM}(\left|r_2\right|,k_{F_2}a)\right]\, ,
\end{eqnarray}
where $v_i=V_i/V$, $n_{H_i}$ and $n_{L_i}$ are, respectively, the fractional volumes and the heavy-atom and light-atom densities of the two domains, which are labeled by the index $i=1,2$.  These quantities are constrained by the conservation of the heavy- and light-atom global densities, which can be written as $v_1n_{H_1}+v_2n_{H_2}=n_H$ and $v_1n_{L_1}+v_2n_{L_2}=n_L$, and by the volume conservation $v_1 +v_2 = 1$. The energy of each domain is given by the equation of state derived in Section~\ref{methods}, and depends on the local interaction parameter $k_{F_i}a=\left[3\pi^2\left(n_{H_i}+n_{L_i}\right)\right]^{1/3}a$, and on the local majority-minority population imbalance, which can be computed as the absolute value of the (local) heavy-light population imbalance $r_i=\left(n_{H_i}-n_{L_i}\right)/(n_{H_i}+n_{L_i})$. The minority-majority mass imbalance $\tilde m$ used in Section~\ref{methods} has to be defined as $\tilde m= m_H/m_L$ if $n_L >n_H$, and $\tilde m = m_L/m_H$ otherwise.
Phase separation takes place when the energy defined in eq.(\ref{EOS_S}) is minimized by two finite domains with different heavy-atom and/or light-atom densities.\\

\begin{figure}[h!]
\epsfxsize=8cm
\epsfbox{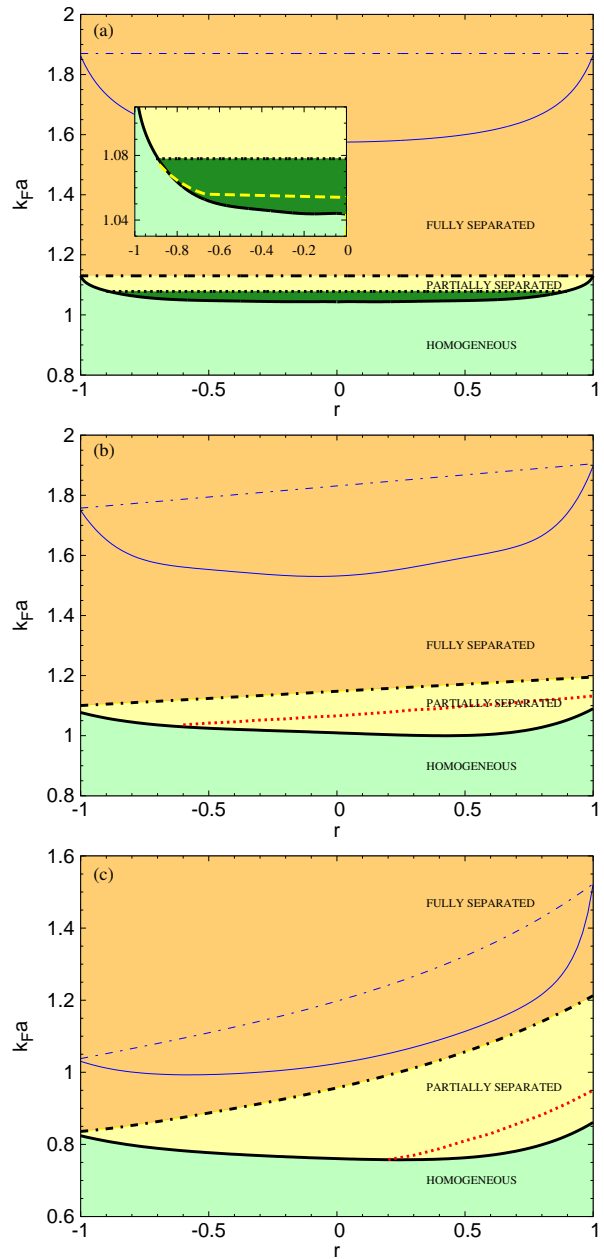}
\caption{Zero-temperature phase diagrams as a function of population imbalance $r=(n_H-n_L)/(n_H+n_L)$ and interaction parameter $k_Fa$, for three mass ratios: $m_L/m_H=1$ (panel (a)), $2/3$ (panel (b)), $6/40$ (panel (c)).  The black continuos and dot-dashed curves indicate the transition to the partially and to the fully-separated states, respectively. The uppermost blue (thin) curves indicate the corresponding mean-field results~\cite{conduit2011}. The dotted red curves separate two kinds of partially-separated states (see text).
The dark-green region in panel (a) indicate the coexistence between the balanced and the partially-separated states. 
The inset in panel (a) is a magnification of the region of phase coexistence. The dashed yellow curve would indicate the direct transition from homogeneous to partially separated~\cite{Duine2005}.} 
\label{phased}
\end{figure}

\begin{figure}[h]
\epsfxsize=8cm
\epsfbox{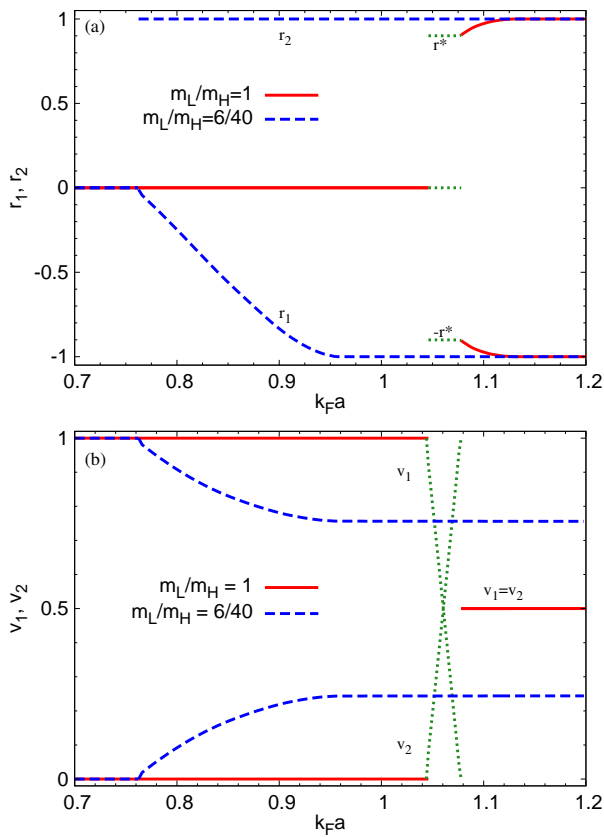}
\caption{Local population imbalances $r_1$ and $r_2$ (panel (a)) and fractional volumes $v_1$ and $v_2$ (panel (b)) of two domains, as a function of the interaction strength $k_Fa$. The populations are globally balanced. The red continuous lines correspond to a mass balanced mixture.The dashed blue lines to a mass imbalance $m_L/m_H=6/40$. The dark-green dotted lines describe the coexistence between a balanced phase in a volume $v_1$ and a partially-separated phase in a volume $v_2$. The subdomains of the partially-separated state have imbalances $\pm r^*$. } 
\label{polvol}
\end{figure}

In Fig.~\ref{phased} we show the phase diagrams in the $r$-$k_Fa$ plane for three different mixtures: a mass-balanced system in panel (a) and two mass-imbalanced configurations, with mass ratios $2/3$ in panel (b) and $6/40$ in panel (c).
The case of panel (c) corresponds to the $^6$Li-$^{40}$K mixture, used in current experiments~\cite{expgrimm}, while the mass ratio $2/3$ (panel (b)), was considered in previous calculations based on mean-field theory.\\
In the last two panels, the lowermost continuos curves divide the homogeneous state (light-green region) from the partially-separated state (yellow region).
In these mass-imbalanced cases, the region of stability of the partially-separated state is crossed by a dotted red curve. This curve divides two kinds of partially-separated states. Between the continuous black curve and the dotted red curve both domains host both heavy and light atoms. Instead, above the dotted red curve (but below the dot-dashed black curve) one domain is fully imbalanced (it contains only one species). One can notice that for global population imbalances $r\lesssim -0.6$ in panel (b) and $r\lesssim 0.2$ in panel (c), the system separates directly by nucleating a fully-imbalanced domain hosting the heavy species only.  
The dot-dashed black curves divide the partially-separated from the fully-separated state (orange region) where both domains host only one species.\\
In the equal-mass case (panel (a)), the population imbalances of the two domains of the separated states have opposite symmetric values. Within second-order perturbation theory, the phase separation is a first-order transition~\cite{Duine2005}, in agreement with the low-energy theory of itinerant fermions of Ref.~\cite{belitz}. Hence, there is a region of coexistence between the balanced phase and the partially-separated state. The region where this coexistence is energetically favourable is indicated in panel (a) with the dark-green color. If the global population imbalance is $r=0$,  the critical interaction parameter $k_Fa \cong 1.046$ is obtained by studying the equilibrium between the balanced phase and a partially-separated state having two symmetric subdomains with equal densities and opposite local population-imbalances ($r_1 = -r_2$). By varying the relative volumes of these two subdomains, a global population imbalance can be accommodated. The maximum global imbalance that can be accommodated is obtained by studying the equilibrium between the balanced phase and a single fully-imbalanced domain. This critical global imbalance is indicated in panel (a) by the continuous black curve. Notice that for global imbalances $|r| \gtrsim 0.9$ the coexistence between the balanced and the partially-separated phases is not possible, so one has a direct transition between these two phases. If we disregard the possibility of phase coexistence also for $|r| \lesssim 0.9$, we obtain a direct transition at the critical interaction indicated by the dashed yellow curve (see inset of panel (a)). This curve converges to $r = 0$ at $k_Fa \cong 1.054$, in agreement with previous studies also based on second-order perturbation theory~\cite{Duine2005}.\\
In Fig. \ref{polvol}, we show the dependence of the local population imbalances in each domain (panel (a)), and the fractional volumes occupied by these domains (panel (b)), as a function of the interaction parameter $k_Fa$. The global populations are balanced ($r=0$). 
%At the beginning, for values of $k_Fa$ smaller than the critical ones, the two mixtures are homogeneous, therefore we can imagine to have only one domain with volume $v_1=1$ and $r_1=0$. By increasing the repulsion, 
In the mixture with mass imbalance $m_L / m_H= 6 /40$ (dashed blue lines), phase separation occurs by nucleating a small domain whose volume (indicated as $v_2$) gradually increases. This domain contains only the heavy atoms, hence its imbalance is $r_2 = 1$. The other domain hosts a majority of light atoms, hence $r_1 < 0$. This imbalance smoothly converges to $r_1 = -1$, where the fully-separated state is reached. Beyond this point the relative volumes of the domains are fixed. The volume of the domain with heavy atoms is less than half of the other domain's volume.
In the mass-balanced mixture, the domains of the separated states are symmetric, hence $r_2 = -r_1$ and $v_1=v_2=0.5$ (because of global population balance). The dashed dark-green lines correspond to the small window where the balanced phase coexists with the partially-separated state. As stated above, the nucleation of the partially-separated state starts at $k_Fa \cong 1.046$. The volume of the nucleated domain gradually increases up to $k_Fa \cong 1.078$. At this point the balanced phase disappears and the partially-separated state with symmetric domains occupies the whole volume. It is worth noticing that in the region of phase coexistence the local population-imbalances of the subdomains of the partially-separated state are fixed at $|r^*| \cong 0.9$.\\
%%%%%%%%%%%%%%%%

%%%%
%%%%%%%%%%%%%%%
In the three panels of Fig.~\ref{phased}, the two uppermost blue (thin) curves correspond to the borders of the phase separated region calculated within the mean-field approximation, as in Refs.~\cite{conduit2011,ho2013}.  
It is evident that introducing beyond mean-field terms in the equation of state determines significant modifications in the phase diagram. The quantum phase transition from the homogeneous to the separated state is strongly favoured compared to the mean-field prediction, in particular in the case of a large majority of heavy atoms. It is worth noticing that, according to the second-order theory, the partially-separated state is stable, in the limit $r \rightarrow 1$, for a finite (and relatively large) range of the interaction parameter. This means that an infinitesimal concentration of light atoms is sufficient to stabilise a separated state where at least one domain hosts both species. Instead, in the mean-field phase diagram the region of stability of the partially-separated state vanishes in the limit of large imbalance.\\

\begin{figure}[h]
\epsfxsize=8cm
\epsfbox{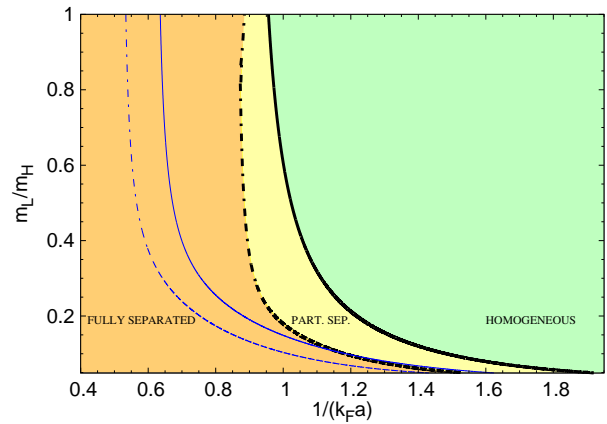}
\caption{Phase diagram as a function of the inverse interaction strength $(k_Fa)^{-1}$ and the light-heavy mass ratio $m_L / m_H$, for a population balanced mixture. 
The continuos black curve represents the boundary between the homogeneous state (light-green region) and the partially-separated state (yellow region). The dot-dashed black curve is the boundary between partially and fully-separated state (orange region). The two leftmost blue (thin) curves indicate the corresponding mean-field results~\cite{conduit2011,ho2013}.}
\label{phased_r0}
\end{figure} 

Fig.~\ref{phased_r0} shows the effect of varying the mass imbalance for globally balanced populations ($r = 0$). We present the phase diagram in the mass-imbalance - inverse interaction parameter plane ($m_L /m_H$-$(k_Fa)^{-1}$). Our results confirm that, at any repulsion strength, it is always possible to reach the separated state by increasing the mass imbalance, as predicted in Ref.~\cite{ho2013}. Our second-order calculation provides critical values of the repulsion strength smaller than the mean-field prediction. We notice that the critical coupling required to reach the fully-separated state has a slightly non-monotonic dependence, with a maximum at an intermediate value of the mass ratio.
We also point out that, in the equal-mass limit $m_L/m_H \rightarrow1$, the line of transition from the homogeneous to the partially-separated state converges to the critical interaction parameter $k_Fa \cong  1.046$, where the phase coexistence between balanced and partially-separated phases takes place (see Fig.~\ref{phased}, panel (a)).

%%%%%%%%%%%%%%%%%%%%%%%%%%%%%%%%%%%% 
%%%%%%%%%%%%%%%%%%%%%%%%%%%%%%%%%% 
 \section{Concluding Remarks}\label{conclusions}
 %%%%%%%%%%%%%%%%%%%%%%%%%%%%%%%%%
In conclusion, we have determined the equation of state of a Fermi-Fermi mixture with both population and mass imbalance using second-order perturbation theory. 
This work extends to the case of unequal masses the previous studies by Lee, Huang and Yang~\cite{huang,lee} and by Kanno~\cite{kanno1,kanno2} who considered, respectively, unpolarized and polarized spin-1/2 Fermi gases.
Making use of this second-order equation of state, we have investigated the zero-temperature phase diagram for varying population imbalance and interaction strength. We observe that increasing the mass imbalance strongly favours phase separation, giving support to the idea of using mixtures of different atomic species as an alternative route to investigate itinerant ferromagnetism with ultracold atoms~\cite{ho2013}. For the equal mass case, our phase diagram agrees with previous findings~\cite{Duine2005} (also obtained within second-order perturbation theory), and we also included the analysis of the phase coexistence at the border of the first order transition.
For the mass ratio corresponding to the $^6$Li-$^{40}$K mixture, the partially-separated state is stable in a relatively large range of interaction strength even in the regime of large majority of heavy atoms.
Compared to previous mean-field studies, our second-order theory predicts considerably weaker critical interaction strengths for phase separation. In the mean-field phase diagram, the region of stability of the partially-separated state vanishes at large population imbalance, contrary to the second-order results. Also, while we find two kinds of partially-separated states, one where each domain contains both species, the other having one fully unbalanced domain containing atoms of one species only, in the mean-field case the partially-separated state is always of the first kind.\\
These findings indicate that atomic Fermi-Fermi mixtures are the ideal experimental setup to investigate the role of beyond mean-field effects in many-fermion systems.
We have provided a detailed analysis of the phase-diagram of the $^6$Li-$^{40}$K mixture, which is indeed relevant for the current experiments performed with ultracold atomic mixtures~\cite{expgrimm}. Repulsive Fermi-Fermi mixtures can also be formed in experiments with one bosonic and one fermionic species~\cite{salomon} by creating fermionic molecules following the attractive branch of a Feshbach resonance. In the case of isotopes, the mass ratio would be $m_L/m_H \simeq 1/2$.
The parametrization of the equation of state provided in Appendix~\ref{appendix} can be used to extend the phase diagrams of Fig.~\ref{phased} to configurations with confinements, relevant for specific experimental setups.\\

 %%%%%%%%%%%%%%%%%%%%%%%%%%%%%%%%%%%%
 We acknowledge useful discussion with G. Bertaina, M. Capone, S. Giorgini and P. Pieri.
 
 %%%
 %%%%%%%%%%%%%%%%%%%%%
 \appendix
 %%%%%%%%%%%%%%%%%%%%%
 \section{Second order contribution in the EOS}
 \label{appendix}
To compute the function $I(r,\tilde m)$ (defined after equation~\ref{Irm1}) we employ a stochastic integration procedure. Below we provide a simple parametrization of the results of the stochastic integration, which we perform on a dense grid of values of the population imbalance $r$, and for various relevant values of the minority-majority mass imbalance $\tilde m=m_</m_>$. We perform a best-fit analysis using the following functional form:
\begin{equation}\label{fit}
I^{\tilde m}(r)=f^{\tilde m}_1(r)(1-f_{damp}(r))+f^{\tilde m}_2(r)f_{damp}(r)\,
\end{equation}
where
\begin{eqnarray}\label{fit}
&&f_{damp}(r)=\frac{1}{2}\left\{\tanh\left[2\pi(r-0.7)\right]+1\right\} , \\\nonumber
&&f^{\tilde m}_1(r)=a_1r^2+b_1r+c_1 , \text{and} \\\nonumber
&&f^{\tilde m}_2(r)=a_2r^2+b_2r-(a_2+b_2).
\end{eqnarray}
%%%%%
%%%%%
In Tab.~\ref{tabfit} we provide the values of the fitting parameters $a_1$, $b_1$, $c_1$, $a_2$ and $b_2$, for the mass ratios $\tilde m = 1/2, 2, 2/3, 3/2, 6/40, 40/6$.
\begin{table}[ptb]
\caption{Coefficients of the fitting functions (\ref{fit}), for several values of the mass ratio $\tilde m$}
\begin{tabular}{c||c|c|c|c|c}\label{tabfit}
$\tilde m$ & $a_1$ & $b_1$ & $c_1$ & $a_2$ & $b_2$ \\ 
\colrule
 $\frac{1}{2}$ &   -1.08798  & 0.135395 & 1.15641 & -2.44043 &  2.06706\\
 2 & -1.95775  &  -0.234915 & 2.3135 & -2.78626 & 0.948627 \\
 $\frac{2}{3}$ & -1.3299  &  0.0996374 & 1.43752 &  -2.70114   &  2.05884\\
 $\frac{3}{2}$ & -1.88724   &   -0.115705  &   2.1563    &   -2.91861  &  1.35634\\
 $\frac{6}{40}$ & -0.327123 & 0.123349  & 0.337514   &  -1.22671   &  1.41427\\
 $\frac{40}{6}$ & -1.51799  &  -0.782685  &  2.25148   &  -1.32853   &   -1.03849
\label{tab:eos}
\end{tabular}
\end{table}


\begin{thebibliography}{99}
\bibitem{noteQMC} A comparison between third-order expansion and quantum Monte Carlo simulations for finite-range potentials was shown in Ref.~\cite{boronat}.
\end{thebibliography}
\end{document}